\documentclass[pdflatex,sn-mathphys-num]{sn-jnl}
\usepackage{url}

\usepackage{graphicx}%
\usepackage{multirow}%
\usepackage{amsmath,amssymb,amsfonts}%
\usepackage{amsthm}%
\usepackage{mathrsfs}%
\usepackage{dsfont}
\usepackage[title]{appendix}%
\usepackage{xcolor}%
\usepackage{textcomp}%
\usepackage{manyfoot}%
\usepackage{booktabs}%
\usepackage{algorithmicx}%
\usepackage{algorithm}
\usepackage{algpseudocode}
\usepackage{listings}%
\usepackage{braket}
\usepackage{tikz}
\definecolor{orcidgreen}{HTML}{A6CE39}
\renewcommand{\orcidlogo}{\begin{tikzpicture}[baseline=-0.55ex]\fill[orcidgreen] (0,0) circle [radius=0.75ex] node[white,inner sep=0pt] {\sffamily\bfseries\fontsize{4.5}{4.5}\selectfont iD};\end{tikzpicture}}

\theoremstyle{thmstyleone}%

\theoremstyle{thmstyletwo}%

\theoremstyle{thmstylethree}%

\begin{document}

    \title[To predict or to persist]{To predict or to persist: how decoherence shapes learning in a spin-qubit quantum reservoir}

\author*[1,2,3]{\fnm{Nikolaos} \sur{Petropoulos}\orcid{https://orcid.org/0000-0001-7831-0082}}\email{nikolaos.petropoulos@equal1.com}

\author[1,2,3]{\fnm{Elena} \sur{Blokhina}\orcid{https://orcid.org/0000-0002-4164-4350}}\email{elena.blokhina@ucd.ie}

\affil[1]{\orgname{Equal1 Labs}, \orgaddress{\street{Nexus UCD}, \postcode{Belfield}, \city{Dublin}, \country{Ireland}}}

\affil[2]{\orgdiv{School of Electrical and Electronics Engineering}, \orgname{University College Dublin}, \orgaddress{\postcode{Belfield}, \city{Dublin}, \country{Ireland}}}

\affil[3]{ \orgname{Centre for Quantum Engineering, Science, and Technology (C-QuEST)}, \orgaddress{\postcode{Belfield}, \city{Dublin}, \country{Ireland}}}

\abstract{ In this work we investigate how dephasing and the choice of measured observables affect learning in a simulated six-qubit spin chain with Markovian dissipation and exact expectation-value readout. For externally driven nonlinear autoregressive moving-average signals, combining single- and two-qubit observables gives better predictions than a population-only readout at comparable polynomial feature counts. Most of the configurations we tested become more accurate than a reference with relaxation and dephasing disabled, most strongly at $T_2=1~\mu s$, where total memory capacity drops, and still at $T_2=5$--$10~\mu s$, where memory is only redistributed across delays. For autonomous generation, we train the readout with scheduled sampling and iterative dataset aggregation to expose it to states reached under its own feedback. We do not find a clear systematic dependence of limit-cycle persistence on coherence across the reservoir ensemble. For chaotic forecasting, in contrast, prediction horizons at $T_2=1~\mu s$ shorten in all 10 reservoir realizations relative to that reference. For the noise-driven strange non-chaotic attractor, the model produces a smooth, approximately closed orbit, and neither captures the target's scatter nor follows its trajectory pointwise. We conclude that, within this simulated reservoir, the impact of coherence depends strongly on the task.}

\keywords{quantum reservoir computing, decoherence, spin qubits, quantum machine learning, time-series prediction, chaotic dynamics}

\maketitle

\section{Introduction}\label{sec:intro}

Predicting a changing signal requires a model to retain useful information about the past and combine it with new observations. Reservoir computing (RC) approaches this problem by using the response of a dynamical system to an input sequence. It exploits the fact that if the system used is complex enough, it performs most of the computation that a recurrent neural network is otherwise required to learn~\cite{Tanaka2019, nakajima_physical_2020, yan_emerging_2024}. After the system is driven with a temporal input, one proceeds by reading out a set of internal state variables and fitting a linear or low-degree polynomial map from those variables to a target sequence. Since only the readout is trained, the training reduces to a convex regression problem. 

If we use a physical dynamical system as a reservoir, it reduces the need to train internal parameters. This has motivated implementation in photonics, spintronics and nanoelectronic materials~\cite{Tanaka2019, Roy2019,Sangwan2020, Lv2020}. Quantum reservoir computing (QRC) is based on the same philosophy, only using a quantum system as a reservoir~\cite{fujii_harnessing_2017, ghosh_quantum_2019-1, mujal_opportunities_2021, fujii_quantum_2020, markovic_quantum_2020}. This ranges from superconducting qubits~\cite{carles2026, suzuki_natural_2022}, photonic~\cite{Garcia2023, Zia2025, Paparelle2026}, nuclear magnetic resonance (NMR) devices~\cite{Negoro2018}, single nonlinear oscillators~\cite{govia_quantum_2021}, analog Rydberg-atom machines~\cite{kornjaca_large-scale_2024, Liu2026}, to interacting spins~\cite{nakajima_boosting_2019, Hou2026} and applications to state tomography and quantum circuit compression~\cite{angelatos_reservoir_2020, ghosh_reconstructing_2021, ghosh_realising_2021}. Recent reviews collect the state of the field~\cite{gyurik_quantum_2026, tariq_quantum_2026}. 
Since the quantum state space is exponentially large in the number of qubits, we can in principle use a small number of physical degrees of freedom to get a large number of nonlinear functions of the input history. The role of a classical neural network is therefore substituted by the reservoir's Hilbert space~\cite{kalfus_hilbert_2022, ghosh_universal_2020, martinez-pena_information_2020}. 

An open debate in the field is whether decoherence helps or hinders learning, and it is important for any potential use of real hardware as a quantum (reservoir) computer. So far in the literature there are arguments on both sides. On the one hand, the echo state property requires the decay of the influence of the past (i.e., finite memory). This favors (quantum) dissipation as a natural way for a quantum system to forget about its previous states and make the map well defined~\cite{sannia_dissipation_2024, monzani_non-unital_2024}. A recent simulation of a six-qubit Heisenberg-coupled spin-qubit array found that larger decay rates improved performance on NARMA tasks up to order $n=10$, because they stopped the reservoir retaining too much of its input history~\cite{mifune_effects_2025}. On that note, ref.~\cite{ricci_quantum_2026} proposes controllable damping as a design parameter of a system. On the other hand, coherence generates the interference between the many nonlinear functions that the Hilbert space supplies in the first place. A many-body study reports a monotonic relationship between coherence and information processing capacity in a transverse-field Ising reservoir, and finds that finite measurement resources and noise limit how far quantum effects can be exploited~\cite{palacios_role_2024}.

We believe that the disagreement is partly a matter of the specific task performed. Specifically, if a reservoir is required to reproduce a nonlinear filter of its input then it needs memory. If it is required to sustain a limit cycle autonomously it needs to have learned an attracting orbit; such orbits usually absorb small perturbations by construction. Conversely, if it is required to track a chaotic trajectory, Lyapunov divergence amplifies the perturbations. We find no reason for three such different demands on the same hardware to respond to decoherence in a similar fashion. In order to test whether coherence dependence changes with the task within a single model, we fix the physical parameters to a realistic operating point and sweep the coherence times across a range.

\begin{figure}[t]
    \centering
    \includegraphics[width=\linewidth]{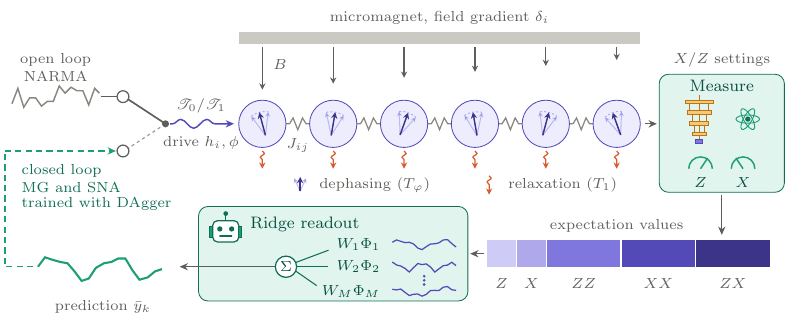}
    \caption{Schematic of the QRC protocol. The input $u_k$ controls the mixture of two completely positive trace-preserving (CPTP) maps $\mathscr{T}_0$ and $\mathscr{T}_1$, which differ in the phase $\phi$ and amplitudes $h_i$ of the transverse drive, in exchange strength and in duration ($2\tau$ and $3\tau$). They act on a six-qubit chain with Heisenberg exchange $J_{ij}$ and a micromagnet field gradient that sets the detunings $\delta_i$. Every qubit relaxes with time $T_1=100$~ms and undergoes pure dephasing with time $T_\varphi$, where $1/T_\varphi=1/T_2-1/(2T_1)$ and we sweep the coherence time $T_2$. Five local $X/Z$ measurement settings give 57 expectation values, and the ridge readout forms the prediction as a weighted sum of $M$ polynomial features $\Phi_j$ of these values. Only the weights $W_j$ are trained. The input switch defines the driving mode: for NARMA the reservoir is driven by an external signal (open loop), while for MG and the SNA it is driven by its own prediction (closed loop), and the readout is trained in the loop with DAgger/scheduled sampling.}
    \label{fig:scheme}
\end{figure}

The reservoir we simulate here is a chain of six electron-spin qubits of the kind realized in Si/SiGe heterostructures, coupled by Heisenberg exchange and addressed through the field gradient of a micromagnet. This is a platform that is able to run multi-qubit circuits in practice~\cite{Vandersypen2019, philips_universal_2022, fernandezdefuentes_running_2026}, and experiments have directly characterized six-dot arrays of this kind. Their relaxation times are of the order $T_1\sim 100~$ms and free-evolution coherence times of order $T_2\sim 10~\mu$s at mK temperatures~\cite{amitonov_persistence_2026, borjans_single-spin_2019, wang_pursuing_2024}. Here we hold $T_1=100~$ms and sweep $T_2$ from $1$--$5000~\mu$s, together with a unitary evolution reference. The parameters we use for the configuration of the reservoir are representative of the platform and not a specific chip. We drive the reservoir stochastically with a non-zero reset probability~\cite{chen_temporal_2020}. Moreover, alongside the coherence sweep we test different sets of readout quantum observables used as learning features. These range from the six single-qubit populations $\langle Z_i \rangle$ up to a set of 57 single-qubit and two-qubit expectation values spanning local $X/Z$ measurements arranged in five settings. We benchmark on three temporal tasks with three different requirements sketched above: i) the $n$-order nonlinear auto-regressive moving average (NARMA-$n$) family, driven open-loop, ii) the Mackey--Glass (MG) system~\cite{Mackey1977} generated autonomously, in both its non-chaotic and chaotic regimes, and iii) a noise-induced strange non-chaotic attractor (SNA)~\cite{Grebogi1984, jin-tu_strange_2008}, which is also autonomous. Figure~\ref{fig:scheme} summarizes the scheme.

Autonomous generation requires a training procedure that is not necessarily included in standard QRC. Fitting the readout on teacher-forced states minimizes one-step error but places no constraint on the stability of the resulting closed-loop map. When the loop is closed, the trajectory can collapse onto a fixed point or drift away from the attractor. This is the same exposure-bias failure that is found in autoregressive sequence models~\cite{ranzato_sequence_2016, he_exposure_2021}. In order to mitigate it, we use \textit{scheduled sampling}~\cite{bengio_scheduled_2015} combined with \textit{dataset aggregation} in the manner of DAgger~\cite{ross_reduction_2011}, so that we fit the readout on the off-attractor states reached by its own predictions, labeled with the true continuation (closed loop in Fig.~\ref{fig:scheme}). We will show that on the MG tasks the in-loop fit is almost always selected over the teacher-forced one. Under the assumptions of our variance analysis, the reservoir realization is estimated to contribute more to performance variability than training randomness.

Here we investigate a device-grounded but still idealized reservoir, since we simulate the Lindblad dynamics and read out exact expectation values. This means that finite measurement statistics, readout infidelity, gate calibration error, and non-Markovian noise are not taken into account. The first one interacts with coherence~\cite{palacios_role_2024} and we return to it in Sec.~\ref{sec:discussion}. Readout infidelity depends on the charge sensor that detects the spin-to-charge conversion, such as a single-electron box~\cite{petropoulos_nanoscale_2024}. We also treat the exchange couplings as a quenched disorder rather than as a controllable resource. This allows us to isolate the effect of coherence, and establish a baseline which could be employed by future work with a realistic measurement overhead.

The structure of the paper is as follows. In Sec.~\ref{sec:results} we present the results for all three tasks, covering what the readout observables buy and then the coherence dependence of each task. Also, we separate the contribution of the reservoir configuration from that of the training stochasticity. In Sec.~\ref{sec:discussion} we compare the findings across tasks and close with our conclusions and remarks. Finally, in Sec.~\ref{sec:methods} we present the reservoir Hamiltonian and its Lindbladian used, the observable sets and the cost of the measurement settings for each of them, and the training and evaluation methodology, including the in-loop algorithm that makes autonomous generation possible.

\section{Results}
\label{sec:results}

We test our quantum reservoir on three separate classes of timeseries, each one with different instantiations. The first one is the nonlinear autoregressive moving-average time series of order $n$ (NARMA-$n$). The second one is the MG timeseries, with lags $\tau_{\textrm{MG}}=16,17$. The two different lags give rise to non-chaotic and chaotic dynamics, respectively. Finally, we attempt to predict a timeseries generated by the noise-driven map of Eq.~\eqref{eq:sna}, whose trajectory in the $(y_k,y_{k-1})$ plane traces an SNA.

We sweep two axes throughout our calculations. One is coherence time $T_2$ from $1~\mu$s to $5$~ms at fixed $T_1=100~$ms, with a unitary reference in which the relaxation and dephasing are switched off in Eq.~\eqref{eq:lind}. The other is the quantum observable set $\mathbb{O}$ measured from the qubits, from 6 up to the full 57 of Eq.~\eqref{eq:omax}. Moreover, all open-loop results are reported for two independent reservoir instantiations, which we label $\mathscr{R}_1$ and $\mathscr{R}_2$. We show in Section~\ref{sec:stochasticity} that the choice of the reservoir is important and strongly affects the results.

\begin{table*}[t]
    \centering
    \caption{Observable readout at equal number of regressor terms. NMSE on NARMA-$n$, median over the nine coherence columns of the sweep, including the unitary reference. $\mathbb{O}_{\mathrm{all}}$ at degree $d=2$ gives 1710 features and $\mathbb{O}_Z$ at $d=7$ gives 1715. The last column counts the cells in which the full set wins, out of 3 orders $\times$ 9 coherence values. }
    \begin{tabular}{llccccc}
    \toprule
    & & \multicolumn{3}{c}{median NMSE} & & \\
    \cmidrule(lr){3-5}
    reservoir & readout & $n=10$ & $n=15$ & $n=20$ & \# features & wins\\
    \midrule
    $\mathscr{R}_1$ & $\mathbb{O}_Z$, $d=7$  & 0.632 & 0.679 & 0.740 & 1715 & --- \\
    $\mathscr{R}_1$ & $\mathbb{O}_{\mathrm{all}}$, $d=2$ & 0.305 & 0.450 & 0.558 & 1710 & 27/27\\
    $\mathscr{R}_2$ & $\mathbb{O}_Z$, $d=7$ & 0.653 & 0.690 & 0.734 & 1715 & --- \\
    $\mathscr{R}_2$ & $\mathbb{O}_{\mathrm{all}}$, $d=2$ & 0.263 & 0.481 & 0.576 & 1710 & 27/27 \\
    \botrule
    \end{tabular}
    \label{tab:obs}
\end{table*}

\subsection{Readout richness at matched feature count}
\label{sec:features}

\begin{figure}[t]
    \centering
    \hspace*{-.8cm}
    \includegraphics[width=1.1\linewidth]{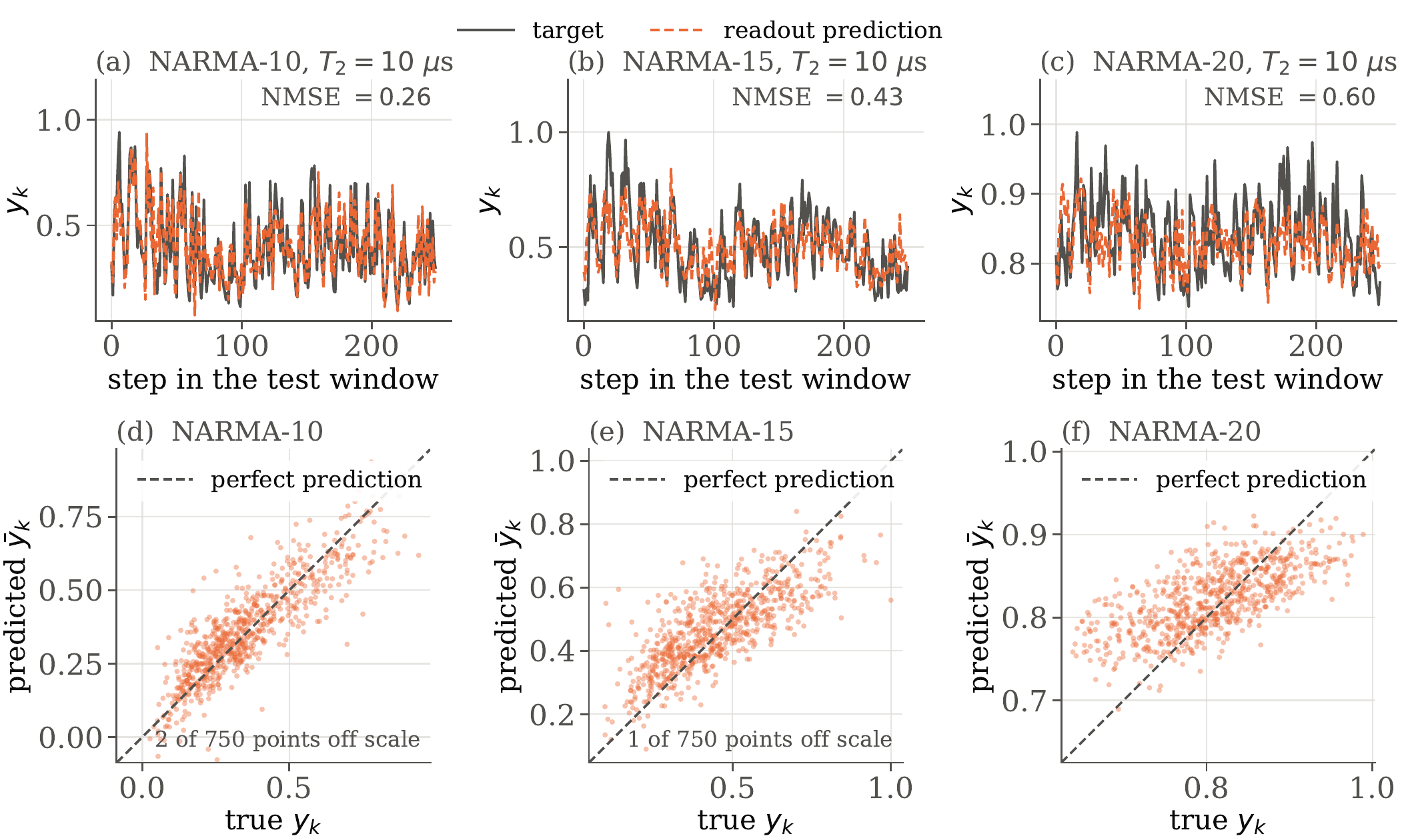}
    \caption{NARMA-$n$ timeseries prediction for orders $n=\{10,15,20\}$. We plot the full observable set $\mathbb{O}_{\mathrm{all}}$ for $\mathrm{deg}=2$ on $\mathscr{R}_1$, at $T_2=10~\mu$s on the three NARMA orders (columns). \textit{Top row:} 250 out of the 750-step test window, target vs prediction, with the NMSE annotated. \textit{Bottom row:} target vs prediction over the whole test window, on equal axes. We count the number of points that fall off the target axis limits. The prediction accuracy degrades with increasing orders as the memory demand grows. We can see (bottom row, from left to right) that the point cloud progressively scatters away from the diagonal, over-predicting smaller values while under-predicting larger ones. At $n=20$ in (f) the reservoir cannot predict the points beyond the center accurately. We highlight that $T_2=10~\mu$s is the realistic device column instead of the best cell. Table~\ref{tab:obs} shows the medians over the whole coherence axis. $\mathscr{R}_1$ was also chosen for consistency, since from the same Table we see that $\mathscr{R}_2$ performs better on one of the three orders $n$.}
    \label{fig:timeseries}
\end{figure}

We investigate whether the extra quantum observables offer any essential improvement in learning by themselves, or they only increase the number of regressors. We compare the full observable set at quadratic order generating 1710 features, against a $Z$-only readout of degree $\mathrm{deg}=7$ generating 1715 features. The two generate approximately the same number of features, to ensure a fair comparison in the regressor count and to distill the information content of them. We show the comparison in Table~\ref{tab:obs}. We can see that the full set wins in all 27 cells of the grid on both reservoirs. We show the readout prediction in Fig.~\ref{fig:timeseries} at $T_2=10~\mu$s, the lower end of free-evolution coherence in Si/SiGe devices.

Given that it is not efficient to have 1715 regressors on six variables, we also ran the $Z$-only readout at $\mathrm{deg}=6$ which yields 923 features. Across the 54 matched cells, the median absolute NMSE difference between the $\mathrm{deg}=6$ and $\mathrm{deg}=7$ readouts is $8\times10^{-5}$, with a maximum of approximately $1.05\times10^{-2}$. This supports the idea that the improvement comes from the additional observables themselves and not from the number of monomials that participate in a regressor. Splitting the additional observables apart, the two-body correlators alone ($\mathbb{O}_{\mathrm{2q}}$, 1080 features) beat the $Z$-only readout at $\mathrm{deg}=7$ in all 27 cells on both reservoirs despite having fewer features. The single-qubit $X/Z$ readout ($\mathbb{O}_{\mathrm{1q}}$, 1819 features) improves on the $\mathrm{deg}=7$ $Z$-only readout in 25/27 cells on $\mathscr{R}_2$. On $\mathscr{R}_1$, its median NMSE across the nine coherence settings differs from the $\mathrm{deg}=7$ $Z$-only baseline by less than 0.002 for each NARMA order. Since $\mathbb{O}_{\mathrm{2q}}$ also contains the 15 $\langle Z_aZ_b\rangle$, which commute with the populations, these sets do not separate the role of non-commutativity from that of two-qubit correlations. The one-setting readout of Sec.~\ref{subsec:cost}, containing the single-qubit $Z$ observables and the 15 $\langle Z_aZ_b\rangle$ correlators, would test how much improvement can be obtained from commuting observables alone. This would provide a useful control, although it would not by itself establish non-commutativity as the cause of any additional benefit from the full observable set. As we show in Section~\ref{subsec:cost}, these extra quantum features also come at a measurement cost.

\subsection{\texorpdfstring{NARMA-$n$}{NARMA-n}: the effect of dephasing on memory}

\label{sec:narma-memory}

\begin{figure}[t]
    \centering
    \hspace*{-1cm}
    \includegraphics[width=1.15\linewidth]{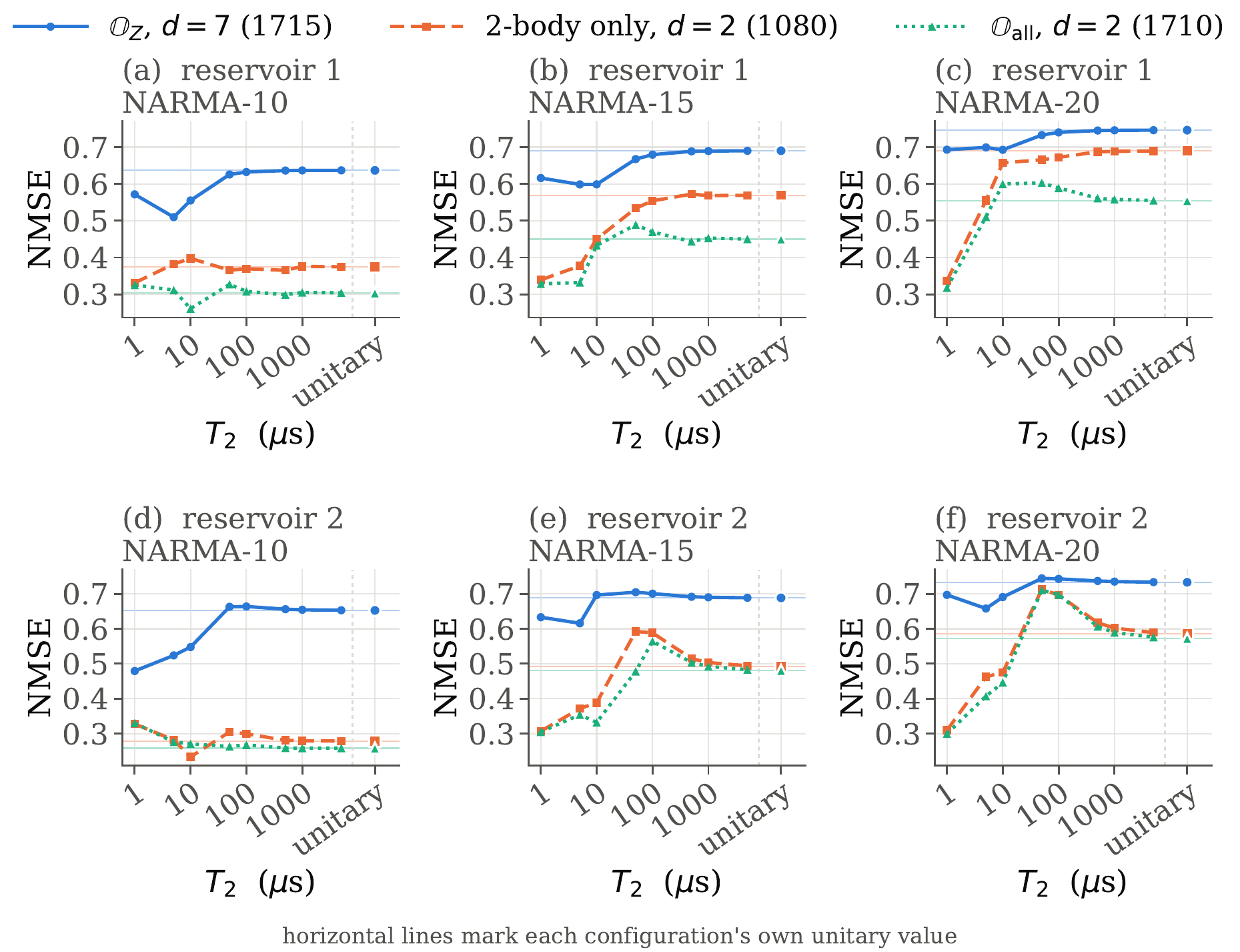}
    \caption{Open-loop NARMA prediction accuracy against coherence time. The rows correspond to the two reservoir instantiations $\mathscr{R}_1$ and $\mathscr{R}_2$ and columns to the orders $n$. We show three observables, $Z$-only at $\mathrm{deg}=7$, two-body correlators at $\mathrm{deg}=2$ and the full set $\mathbb{O}_{\mathrm{all}}$ at $\mathrm{deg}=2$. Each one has 1715, 1080 and 1710 regressors, respectively. The unitary value of each configuration is shown as a projected line with a marker. We can see that the $\mathbb{O}_Z$ readout is worse compared to the other two over the whole range and that the correlated readouts generally improve as $T_2$ decreases. On $\mathscr{R}_2$ their NMSE rises above the unitary value near $T_2=50$--$100~\mu$s and returns to it by $1000~\mu$s.}
    \label{fig:nmse}
\end{figure}

To study the effect of quantum coherence, we pool two reservoirs and five observable configurations over the three NARMA orders. This gives us a total of 30 paired comparisons against the unitary reference at each $T_2$. Since two of the five configurations are the same measurement at two polynomial degrees, excluding one leaves 24 paired comparisons. We can see in Fig.~\ref{fig:nmse} and Table~\ref{tab:narma}, in the lower $T_2$ regime, that dephasing \textit{improves} NARMA performance, and there is a gradual improvement. At $T_2=1~\mu$s the reservoir outperforms its coherent version in 27/30 pairs with a median NMSE ratio of 0.786, at 5~$\mu$s in 24/30, at 10~$\mu$s in 23/30, and for $\geq 50~\mu$s the median NMSE ratio approaches unity. 

\begin{figure}[t]
    \centering
    \hspace*{-.5cm}
    \includegraphics[width=1.05\linewidth]{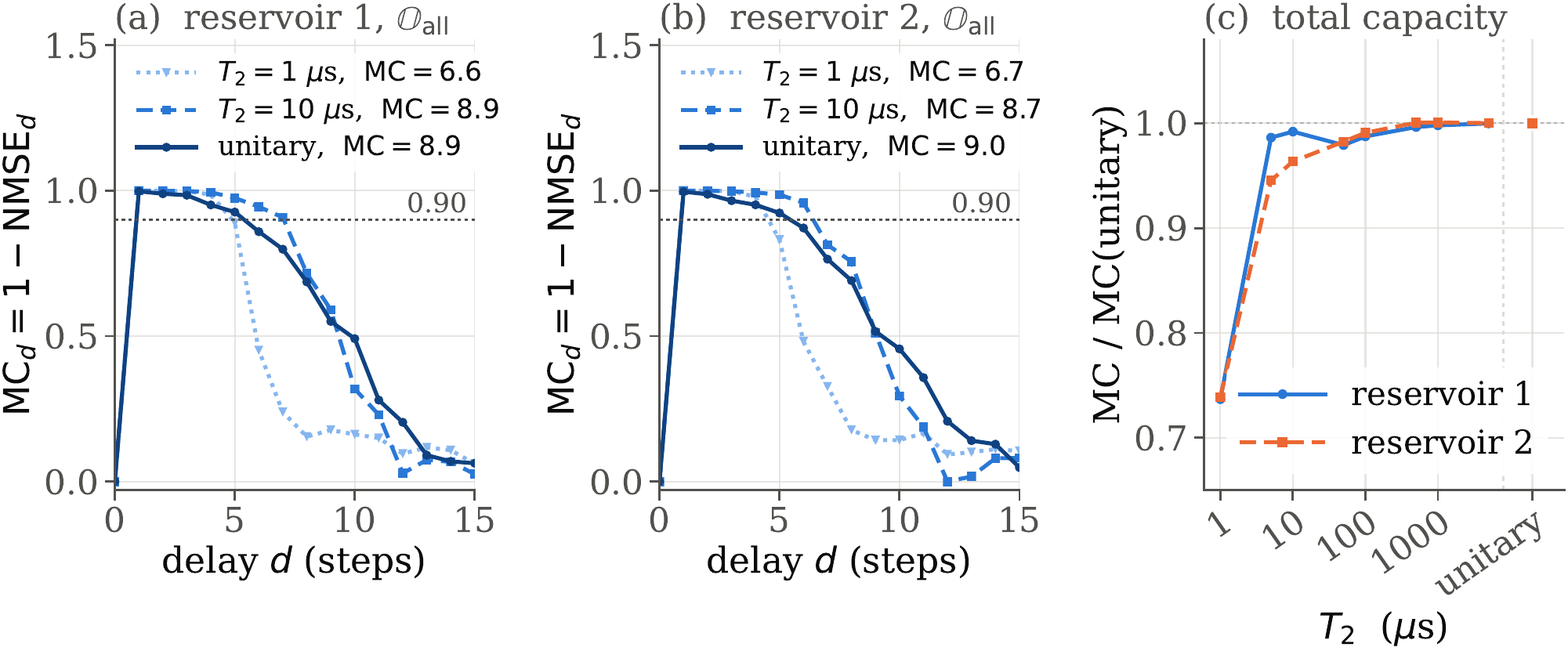}
    \caption{Linear MC of the $\mathbb{O}_{\mathrm{all}}$ readout set. (a,b) Per-delay profile $\mathrm{MC}_d$ for $\mathscr{R}_1$ and $\mathscr{R}_2$ at three coherence values. $\mathrm{MC}_0=0$ by construction, since $\rho_k$ has not yet received $u_k$ (Eq.~\eqref{eq:evo}). Each legend shows its curve's total $\mathrm{MC}$ capacity, where the dotted lines show the 0.90 level used in the text. At $T_2=10~\mu$s the curves lie above the unitary one at every delay up to $d=9$ on $\mathscr{R}_1$ and $d=8$ on $\mathscr{R}_2$, and then drop below it. (c) Total capacity relative to the unitary, showing that $5$--$10~\mu$s columns lose a small amount of memory, while the $1~\mu$s column loses approximately $26\%$.}
    \label{fig:mc}
\end{figure}

At first sight, our results appear inconsistent with the view that nonlinear filtering of an external input requires memory depth. If task performance were controlled primarily by memory, then shortening that memory should reduce accuracy. We investigate this apparent discrepancy by measuring the memory capacity (MC). Using standard construction, we fit each delay $u_{k-d}$ from the feature vector and record $\mathrm{MC}_d = 1 - \mathrm{NMSE}_d$. The total capacity is $\mathrm{MC}=\sum_d \mathrm{MC}_d$. As we can see in Fig.~\ref{fig:mc}, at $T_2=5$ and $10~\mu$s the total capacity is between 0.95 and 0.99 of the unitary value. This shows that essentially no memory has been lost and what changes is its profile. The deepest delay still recovered above 0.90 extends from $d=4$ to $d=5$ steps in the unitary reservoir, to $d=6$ and $d=7$ under dephasing. This is true on both reservoir realizations and correlator configurations. The long weak tail is traded for a sharper cutoff: recall stays above 0.90 to longer delays and then falls faster. This does not improve linear recall of $u_{k-n}$, which is the far tap of the product $u_{k-n}u_{k-1}$ through which the input enters the recurrence for $y_k$. $\mathrm{MC}_{10}$ falls from 0.49 and 0.46 in the unitary reservoir to between 0.26 and 0.33 at $T_2 = 5$ -- $10~\mu$s, and $\mathrm{MC}_{15}$ stays below 0.1 throughout, so the accuracy gain is not explained by better recall of $u_{k-n}$. We return to this below.

\begin{table*}[t]
\centering
\caption{Coherence dependence of the NARMA timeseries. The statistics are calculated over two reservoirs, five observable configurations (i.e $\mathbb{O}_Z$ at $d=6$ and $d=7$, $\mathbb{O}_{\mathrm{1q}}$ at $d=4$, and $\mathbb{O}_{\mathrm{2q}}$ and $\mathbb{O}_{\mathrm{all}}$ at $d=2$) and three orders. In total, we have $30$ paired comparisons against the unitary reference per row. A ratio below unity means the dephased reservoir is more accurate. Two of the five configurations, $\mathbb{O}_Z$ at degrees six and seven, are the same measurement and are not independent; dropping one of them leaves $24$ pairs.}
\label{tab:narma}
\begin{tabular}{lcc}
\toprule
$T_2~(\mu$s$)$ & better than unitary & median NMSE ratio\\
\midrule
1    & 27/30 & 0.786  \\
5    & 24/30 & 0.879  \\
10   & 23/30 & 0.900  \\
50   & 12/30 & 1.016 \\
100  & 11/30 & 1.015  \\
500  & 12/30 & 1.003  \\
1000 & 11/30 & 1.003  \\
5000 & 10/30 & 1.000  \\
\botrule
\end{tabular}
\end{table*}

In order to calibrate the requirements of the tasks, we fit Eq.~\eqref{eq:tap} at increasing tap depth. The error on NARMA-$n$ falls quickly once $\kappa$ exceeds $n$, at $\kappa=11,18$ and 21 for the three orders and is controlled by the tap count. The polynomial degree does not seem to affect it since on NARMA-10 a cubic in 10 taps does not show any improvement over a quadratic in 10, while a quadratic of 11 improves by a factor of 4.3. Moreover, on higher orders the same step from 10 to 11 taps does not change anything, since 11 is still short for their requirements. Therefore, NARMA-$n$ demands close to $n+1$ steps of input memory. The reservoir does not have that, since the deepest delay it recovers above 0.90 is 7, and then only for the $\mathbb{O}_{\mathrm{all}}$ set on $\mathscr{R}_1$ at $T_2=5$ and 10 $\mu$s. Moreover, it is 6 for the same set on $\mathscr{R}_2$, and 6 for the correlator-only readout on both. For the $\mathbb{O}_Z$ set it is at most 2, and on $\mathscr{R}_1$ this set does not reach 0.90 at any delay in six of its nine columns. We see that none of the three orders is being solved with the memory it requires, and the higher two are being solved with less. This is why the coherence dependence tells us more about how a short memory is shaped and less about how far back it reaches. On the other hand, the selected ridge parameter $\alpha_r$ tells the same story from the other side, since the median $\log_{10}\alpha_r$ chosen by cross-validation rises monotonically from $-6.0$ at $T_2=1~\mu$s to $-2.75$ at unitary. This shows that the dephased feature sets need less regularization, which is what we would expect if dephasing had removed redundant directions. We note that $-6.0$ is the floor of our $\alpha_r$ grid rather than an interior optimum. Specifically, at $T_2=1~\mu$s the selection remains on that minimum in 15/15 cells on $\mathscr{R}_1$ and 13/15 on $\mathscr{R}_2$, leaving open the possibility that smaller penalties would further improve validation performance.

At $T_2=1~\mu$s the picture is different and we do not have a sufficient explanation for it. The total capacity falls to 0.74 of unitary on both reservoirs and recall at intermediate delays is reduced. $\mathrm{MC}_d$ is 0.89, 0.45, 0.24, 0.15 on $\mathscr{R}_1$ for $d=\{5, \ldots ,8\}$ against the unitary 0.93, 0.86, 0.80, 0.69, and 0.83, 0.48, 0.33, 0.18 against 0.92, 0.87, 0.76, 0.69 on $\mathscr{R}_2$. Nevertheless, this is the best performing column in the whole sweep. Moreover the NMSE there stops depending on the NARMA order, having 0.33, 0.30, 0.30 values for $n =$ 10, 15, 20 on $\mathscr{R}_2$, while the unitary reservoir degrades from 0.26 to 0.57. The similar errors across the three orders raise the question of which temporal information the predictor uses in this regime. To test this, we checked whether these cells had collapsed into a static map of the recent input, fitting Eq.~\eqref{eq:tap} at tap depths that are too short to succeed in the task. We found that they did not, since a quadratic in 10 taps reached only 0.62 to 0.71 against the 0.30 these cells achieved. We therefore report the accuracy gain at $T_2=1~\mu s$ without proposing a mechanism for it, and note that this column lies an order of magnitude below the typical lower end of coherence on this platform.

\begin{figure}[t]
    \centering
    \hspace*{-1cm}
    \includegraphics[width=1.05\linewidth]{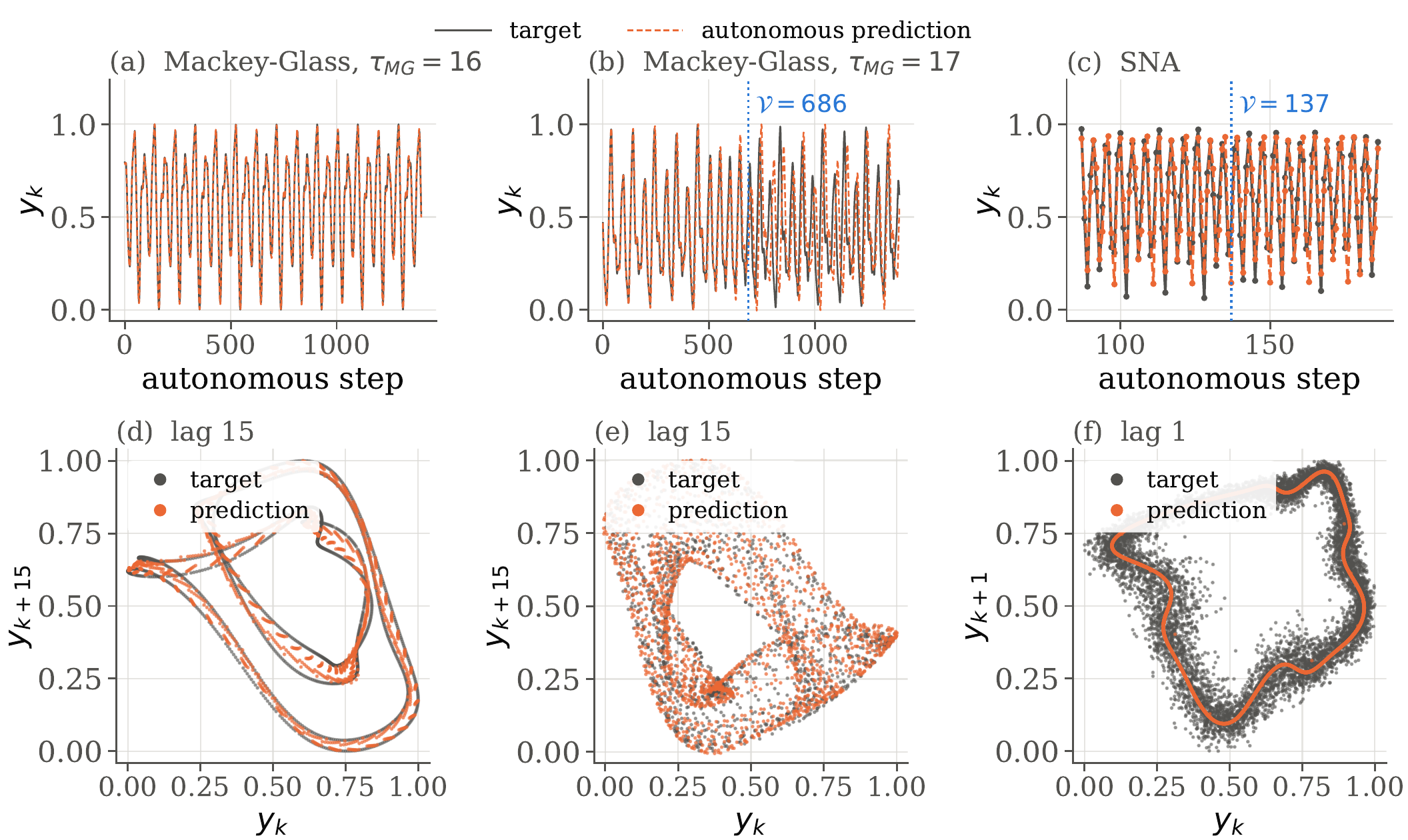}
    \caption{Autonomous phase of the reservoir at $T_2=10~\mu$s. The MG-16 panels show the best of the 10 realizations at this coherence; the MG-17 and SNA panels show $\mathscr{R}_1$, which is also the best realization for MG-17. \textit{Top row:} autonomous trajectory vs truth, with the teacher-forced warmup removed. The valid prediction time $\mathcal{V}$ of Eq.~\eqref{eq:vtime} is marked where it falls inside the window. The MG panels show 1400 steps from the first autonomous one. The SNA panel shows 100 for clarity centered on $\mathcal{V}$. \textit{Bottom row:} the whole autonomous window phase space in delay coordinates, with target and prediction overlaid. Here we test whether the attractor topology was reproduced instead of just tracking. The MG panels use a lag of 15 steps and the SNA of one, since the latter's structure lives in successive steps. In (f) the target occupies a scattered band, while the prediction is a smooth, approximately closed curve in its middle. This non-scattered skeleton behavior is discussed in Sec.~\ref{sec:sna}.}
    \label{fig:mg_sna}
\end{figure}

\subsection{Mackey--Glass: the non-chaotic limit cycle}

We now close the loop and the output is shown in Fig.~\ref{fig:mg_sna}. For $\tau_{MG}=16$ we checked whether the trained readout has the target periodic orbit as an attractor. In our sweep, we ran 70 runs in total which included 10 reservoir realizations and seven coherence columns. We found that the reservoir sustains (i.e., $\mathcal{V}=K_\mathrm{eval}$) all 1699 evaluated autonomous steps in 50 cases, collapsing in approximately $30\%$ of the time. The sustain rate does not seem to depend on coherence since it is essentially flat throughout the $T_2$ sweep, remaining between 6/10 and 8/10 in every column, including the unitary reference.

On the other hand, coherence seems to show up instead in the accuracy of the runs that survive. Fixing the reservoir at $\mathscr{R}_1$ and re-rolling only the training stochasticity, the $T_2=10$ and $50~\mu$s columns beat the unitary reference in all 10 training seeds, with median NMSE ratios of 0.447 and 0.059. Unfortunately, this does not survive changing the chip, since across the 10 reservoir realizations the same two columns win only 5/10 and 7/10, respectively. We flag this explicitly, to avoid overclaiming the beneficial effect of decoherence and to highlight that on one particular structure dephasing helps the non-chaotic task, while on another it does not. This makes the use of dephasing as a resource for this particular task device-dependent.

\subsection{Mackey--Glass: the chaotic regime}

\begin{figure}[t]
    \centering
    \hspace*{-.5cm}
    \includegraphics[width=\linewidth]{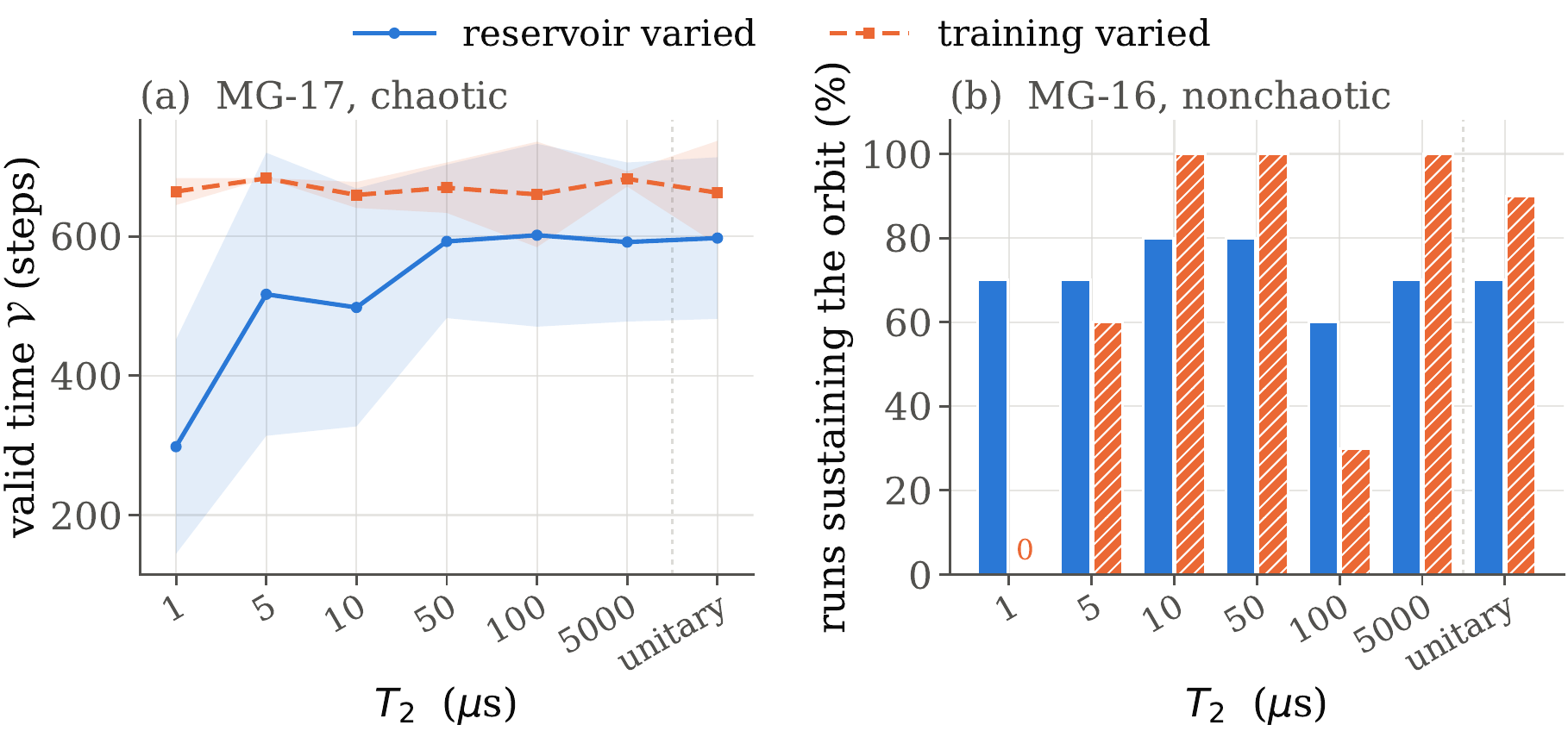}
    \caption{Coherence and hardware effects on autonomous phase MG-$\tau$ generation. (a) Valid prediction time on the chaotic target against coherence. Here we compare i) 10 values of a single seed driving both the couplings and the training RNG, and ii) one reservoir while sweeping 10 training seeds. The bands are 1$\sigma$. The prediction horizon at $T_2=1~\mu$s is shorter than the unitary reference in all 10 reservoir realizations. (b) Fraction of non-chaotic runs that sustain the orbit over all 1699 evaluated autonomous steps. The rate stays between $60\%$ and $80\%$ in every column across reservoirs, while on the fixed chain $\mathscr{R}_1$ it ranges from $0\%$ to $100\%$.}
    \label{fig:vtime}
\end{figure}

At $\tau_{MG}=17$ the attractor becomes chaotic and the more relevant metric becomes the valid prediction time $\mathcal{V}$, defined in Eq.~\eqref{eq:vtime}. In Fig.~\ref{fig:vtime}a, we can see that the effect at $T_2=1~\mu$s is unambiguous and has the opposite sign to NARMA. The unitary reservoir sustains the trajectory for $597.7\pm 116.0$ steps averaged over 10 realizations, and at $T_2=1~\mu$s this falls to $298.2\pm 153.8$. We see the degrading effect in all 10 realizations with a median loss of 313 steps. Among the remaining tested coherence times, $T_2\geq5~\mu$s, none gives a shorter prediction horizon than the unitary reference in more than six of the ten reservoir realizations. Across these values, we observe no consistent increase in prediction horizon with increasing coherence time. The sweep therefore identifies a reduction at $T_2=1~\mu$s but does not establish the precise location of a threshold.

Furthermore, given that the NARMA task was sensitive to the choice of the ridge parameter $\alpha_r$, we checked the same for this task. Specifically, we ran a comparison at $\log_{10}\alpha_r\in\{-6,-4,-2, 0\}$ across three training seeds. The unitary reservoir was better in both $\mathcal{V}$ and NMSE in all 12 combinations with a median NMSE ratio of 2.25, which indicates that the effect is not an artifact of regularization. Moreover, the selected DAgger iteration was always different between the two coherences. The unitary runs peaked, in all combinations, at iterations $13$--$15$ with feedback probability $p_q=0.69$--$0.80$, while the $T_2=1~\mu$s at $9$--$13$ with $p_q=0.48$--$0.69$. Hence, the dephased reservoir exits the scheduled-sampling ramp earlier, which may indicate that its closed-loop map loses stability at a lower feedback probability.

\subsection{Separating reservoir disorder from training stochasticity}

\label{sec:stochasticity}

We swept two randomness axes to quantify the sensitivity to both the reservoir configuration (e.g. exchange $J_{ij}$ disorder), and to stochasticity in the training procedure (e.g. DAgger/ridge training). We compared a mixed sweep with a training-only control, in which the chain was fixed at $\mathscr{R}_1$ and only the training seed was re-sampled over 10 runs. We show the results in Fig.~\ref{fig:vtime}.

Pooled over the 14 cells, re-rolling the seed that drives both the couplings and the training RNG gives a root-mean-square (RMS) spread in $\log_{10}$NMSE 2.12 times that of re-rolling only the training seed, and the spread is wider in every cell. Since the mixed sweep contains both sources, the reservoir contribution can be isolated by subtracting the training variance in quadrature. Here we assume that the training variance is chain-independent and the two sources add in variance, $\sigma^2_{\mathrm{both}}\approx\sigma^2_{\mathrm{chain}}+\sigma^2_{\mathrm{training}}$, with no interaction term. The reservoir contribution is then 1.87 times the training contribution, and it is larger in 10/14 cells, with all four exceptions lying in the chaotic regime.

The distinction between the two is clearer for $\tau_{MG}=17$ at $T_2=5~\mu$s. Across 10 reservoir configurations, the prediction horizon is $517.0 \pm 203.2$ steps, while 10 independent training runs on the fixed reservoir $\mathscr{R}_1$ give $683.4\pm 0.5$ steps. In this cell, the training outcome is reproducible, while the modification of the disorder configuration can shift the prediction horizon substantially.

In addition, the failure statistics point the same way. For the chaotic $\tau_{\mathrm{MG}}=17$ case at $T_2=1\,\mu\mathrm{s}$, we define collapse as a horizon below $70\%$ of the corresponding unitary mean and find that it occurs for $8/10$ reservoir configurations. On the contrary, we do not see a collapse when we fix the chain in any of the ten training seeds. The loss of predictive performance is therefore primarily associated with the reservoir realization. 

For the non-chaotic $\tau_{MG}=16$ case, fixing the chain leaves the pooled collapse rate almost unchanged, at $29\%$ across the mixed sweep and $31\%$ on $\mathscr{R}_1$. Within a column, however, the training seed rarely changes the outcome on $\mathscr{R}_1$. No seed sustains the orbit at $T_2=1~\mu$s, whereas all 10 do at 10, 50 and $5000~\mu$s, and nine do at the unitary reference. The fixed-chain rate therefore records which coherence columns fail on $\mathscr{R}_1$, consistent with the chain term exceeding the training term in all seven non-chaotic cells. On a single chain the limit cycle depends strongly on coherence, and the dependence averages out to between 6/10 and 8/10 per column across reservoirs.

\subsection{The strange non-chaotic attractor}
\label{sec:sna}

The reservoir failed on the SNA by the pointwise measure. We ran the same protocol on 10 reservoir configurations, with $T_1=100$~ms and the eight $T_2$ values of the NARMA sweep plus the unitary reference, giving 90 cells in total. NMSE never falls below 1.471, with a median of 1.834, so the readout does worse than predicting the target mean in all 90 cells. Therefore, the failure is a property of the task on this substrate and does not depend on the realized chain. 

In Fig.~\ref{fig:mg_sna}f, the target occupies a broad scattered band, while the prediction follows a smooth, approximately closed curve through its middle. The prediction captures the broad orbital shape but misses the target's scatter. The noise $\zeta_k$ in Eq.~\eqref{eq:sna} is redrawn independently at each step and is not determined by the state. An autonomous predictor without access to these future noise draws cannot reproduce the target's specific random fluctuations.

However, this is not why NMSE $>1$. Iterating Eq.~\eqref{eq:sna} with $D=0$ gives a period $\mathcal{T}=13$ orbit and its variance is 1.01 times that of the noisy series. The noise-free and noisy series therefore have similar total variances. Applying the exact deterministic map one step ahead to the noisy series gives NMSE $=0.0027$. This means that the failure is most probably due to \textit{phase}, not scatter. A copy of the target displaced along its own orbit by less than a period scores between 0.31 and 3.83 with a mean of 2.17. Our measured range, 1.47 to 1.86, sits inside that band and is expected to be produced by a model reproducing the orbit while drifting in phase. This is also clear from Fig.~\ref{fig:mg_sna}f, where we can see that the geometry survives even though pointwise agreement is not achieved. Hence, the result tests what autonomous generation can achieve on a noise-driven attractor; this perspective is also supported by the separation horizon. It takes two values across the 90 cells, namely $\mathcal{V}=137$ and $\mathcal{V}=202$ in 67/90 and the remaining 23, respectively. Both are in agreement to $7\,\mathrm{mod}\,\mathcal{T}$. More explicitly, $137 = 10 \times 13 + 7$ and $202 = 15 \times 13 + 7$, so the two differ by five loops of the attractor. We expect that a model that captures only the orbit without the phase would produce a horizon that moves in integer periods. In contrast, a model that had lost the attractor at some moment would have no reason to lose it on a lattice. NMSE is bimodal in the same way with 28 cells in $[1.471, 1.650]$ and 62 in $[1.804,1.862]$. Moreover, the two structures agree that 19/23 at $\mathcal{V}=202$ are in the lower cluster against 9/67 at $\mathcal{V}=137$. The teacher-forced bootstrap ($q=0$) is selected in 18/23 cells with the longer horizon and 15/67 with the shorter horizon. These associations suggest that phase alignment and readout selection influence the measured scores, without establishing phase as the only source of error.

\section{Discussion}
\label{sec:discussion}

In this work, we have swept the coherence time $T_2$ of a six-qubit spin chain over a parameter range and analyzed how it affects our three temporal tasks. We observed four different behavioral regimes, since the MG task exhibits different dynamics on each side of its chaotic transition. For the open-loop NARMA family, dephasing helps since the reservoir outperforms its own coherent version in 27/30 paired comparisons at $T_2=1~\mu$s and in 23/30 at $10~\mu$s, while the effect seems to diminish by $50~\mu$s. For autonomous generation of a non-chaotic limit cycle, coherence shows no systematic effect across reservoirs. The failure rate stays between 20 and 40\% per column, including the unitary reference. In contrast, strong dephasing at $T_2=1~\mu$s shortens the chaotic prediction horizon in all ten reservoir realizations. The mean horizon decreases from $598\pm116$ steps in the unitary reference to $298\pm154$ steps at this coherence time. At the other tested coherence times, $T_2\geq5~\mu$s, we observe no consistent increase in prediction horizon with increasing coherence time. Furthermore, for autonomous generation of a noise-driven SNA no configuration at any column of any realization is able to beat the mean prediction. In this task, the variation across cells is consistent with differences in the phase to which the readout locks. In summary, whether decoherence helps or hinders learning in our simulated reservoir depends on the target task.

Our NARMA result agrees in direction with ref.~\cite{mifune_effects_2025}, which found better performance at larger dissipation in a six-qubit Heisenberg-coupled spin-qubit array for NARMA up to order $n=10$; at order $n=20$, where our gain is largest, their larger decay rate is slightly worse. Our memory capacity measurements show that dephasing redistributes the memory profile. At $T_2=5$--$10~\mu$s, the total capacity remains close to the unitary value and high-accuracy recall extends to longer delays. However, there is no improvement on the recall of the distant input $u_{k-n}$. These measurements therefore characterize a change accompanying the accuracy gain, without establishing its mechanism. For chaotic forecasting, the horizon is shorter at $T_2=1~\mu$s, but the sampled coherence values do not establish the precise location of a threshold. Ref.~\cite{palacios_role_2024} reports a monotonic relationship between coherence and information processing capacity in a transverse-field Ising reservoir, so the two results are not directly comparable. Task dependence is one possible explanation for the differing findings, but since we did not reproduce either setting, we cannot establish that explanation.

We highlight two caveats which limit how far this generalizes. First, we read out \textit{exact} expectation values, a point which ref.~\cite{palacios_role_2024} identifies as an important factor. With a finite-shot budget, the estimator variance on each observable competes with the signal. Therefore, an argument that dephasing improves the conditioning of the feature set has to be tested when the features are noisy. Our accounting of the measurement settings in Sec.~\ref{subsec:cost} is a first step in the resource direction. Second, our Lindblad evolution is Markovian, whereas the free-evolution dephasing of a spin qubit is dominated by non-Markovian quasi-static charge and nuclear noise. We expect it to behave differently, since a slowly drifting detuning is closer to a fluctuating reservoir than the decay channel we employed. 

Under the assumptions of chain-independent training variance and additive variance contributions stated in Sec.~\ref{sec:stochasticity}, we estimate that the reservoir component of the RMS spread of $\log_{10}$NMSE is 1.87 times the training component. For example, at $\tau_{MG}=17$ and $T_2=5~\mu$s, 10 realizations spread the horizon over $517.0\pm 203.2$ steps, while 10 training seeds on one reservoir give $683.4 \pm 0.5$. The training protocol can be highly reproducible on a given chain, while the reservoir configuration remains a key factor in learning performance. This makes the coupling pattern itself a design parameter. The disorder strength alone moves a spin reservoir between localized and ergodic dynamics, with the best MG prediction found near the transition~\cite{xia_reservoir_2022}. Beyond disorder strength, structured patterns such as dimerized chains of Su--Schrieffer--Heeger type~\cite{Petropoulos2022} or couplings engineered for perfect state transfer~\cite{christandl_perfect_2004} are possible alternatives to the random draws used here.

In conclusion, we have shown that readout quantum observables contribute information non-trivially and that this is not due to the fact that they add more regressors. This contribution comes at the cost of five measurement settings, which is minimal for six qubits, while a $Z$-basis readout needs one. Also, the aggregated fit was selected over the teacher-forced bootstrap in all 70 non-chaotic runs of the coherence sweep, although on the SNA the bootstrap was selected in 33/90 cells.

We studied NARMA-$n$ timeseries, which are solved efficiently by classical delay-embedding and next-generation reservoir methods~\cite{gauthier_next_2021, bollt_explaining_2021}. Here, we used them as a reference of how the reservoir's behavior is changed by its coherence and readout. For the autonomous tasks we did not run the equivalent delay-line comparison, although next-generation reservoir computing also forecasts autonomously from a delay embedding~\cite{gauthier_next_2021}, which makes it the natural baseline for future work.

For future research, we aim to put a shot budget on the readout and repeat the coherence sweep with sampled expectation values instead of the exact ones that we used here. We believe that this is the assumption most likely to change the open-loop conclusion. A second extension worth investigating is time multiplexing~\cite{fujii_harnessing_2017}. In that case, the reservoir is sampled multiple times within each input interval and its feature vector is built from those sub-steps. With it, a seven-qubit reservoir matched the NARMA performance of a 500-node echo state network. In our implementation, the feature vector was functional instead of temporal, where we expanded 57 observables measured once per step into a polynomial basis. Sub-step sampling requires no additional measurement bases, but on hardware each sub-step needs its own shots, so which of the two approaches is cheaper depends on the platform. It would also be worth replacing the Markovian dephasing with a quasi-static detuning drift, to make the noise model more realistic for spin-qubit devices. The reset channel of Eq.~\eqref{eq:evo}, which gives the reservoir its fading memory, could also be realized physically with metallic leads that initialize the chain by injecting electrons. In our previous work, we have modeled such leads with Anderson impurity models, and in a quantum dot array they changed quantum information scrambling dynamics~\cite{petropoulos_impact_2025}, so they may also change what the reservoir retains. Lastly, we expect that a model with access to its own driving noise instead of a purely deterministic feedback map could potentially recover the SNA's scatter, in addition to its deterministic orbit. That way we could determine whether the limit we report here is the attractor's or the reservoir's.

\section{Methods}\label{sec:methods}

\subsection{The spin-chain reservoir}

\subsubsection{Hamiltonian}

We use a linear chain of $N=6$ electron spin qubits confined in (gate-defined) quantum dots, in a way similar to Si/SiGe devices. Neighboring spins interact through the exchange interaction, which in the single-band Hubbard description reduces to an isotropic Heisenberg coupling in the strong Coulomb interaction regime. Each qubit has its own Zeeman frequency because of the stray field gradient of an on-chip micromagnet for individual addressability. Writing the reservoir Hamiltonian in the rotating frame and ordinary frequency units, we have:
\begin{equation}
    H = \sum_{i=1}^N \delta_i Z_i + \sum_{\langle i,j\rangle} J_{ij} \Big(X_i X_j + Y_i Y_j + Z_i Z_j\Big) + \sum_{i=1}^N h_i\, \Big(\cos \phi\, X_i - \sin \phi \,Y_i\Big)
    \label{eq:ham}
\end{equation}
where $X_i, Y_i, Z_i$ are the single qubit Pauli operators acting on qubit $i$, $\delta_i$ is the detuning of qubit $i$ from the rotating-frame reference, $J_{ij}$ the exchange coupling on the bond $\langle i,j \rangle$ where $\langle \cdots\rangle$ denotes nearest-neighbor indexing, $h_i$ the amplitude of the transverse drive on qubit $i$ and $\phi$ its phase in the XY plane. Eq.~\eqref{eq:ham} is written in frequency units so the time evolution operator is $U(t)=\exp(-2\pi i H t)$. In this normalization, qubit $i$ precesses at $2\delta_i$ in the rotating frame, the drive produces Rabi oscillations at $2h_i$, and the exchange splitting of bond $\langle i,j\rangle$ is $4J_{ij}$.

The detunings created by the micromagnet gradient are taken as $\boldsymbol{\delta}=(28, 14,0,-14,-28,-42)$~MHz. This results in a nearest-neighbor spacing of $14$~MHz and a total spread of $70$~MHz across the $N=6$ dots. The drive amplitudes are $h_i\approx1$~MHz, i.e., Rabi frequencies of about 2~MHz. The exchange couplings with induced disorder are drawn from base values of $(8,5,9,6,7)$~MHz on the five bonds of the chain, where each one is multiplied by an independent factor sampled from $\mathcal{U}(1,2)$. The final couplings are in the range between $5$--$18$~MHz. The disorder is quenched and every result below at a given instantiation of the quantum dot array uses the same chain. The two realizations carried through the open-loop results as $\mathscr{R}_1$ and $\mathscr{R}_2$ have couplings $(14.19, 7.19, 16.73, 10.18, 7.66)$ and $(13.22, 5.22, 9.18, 11.04, 11.11)$~MHz, respectively. We show in Section~\ref{sec:stochasticity} that the reservoir configuration matters significantly.

\subsubsection{Dissipative dynamics}
We treat the reservoir as an open quantum system. Each qubit is subject to amplitude damping at a rate of $1/T_1$ and to pure dephasing at a rate $1/T_\varphi = 1/T_2-1/(2T_1)$, with $T_2\leq 2T_1$. The Lindbladian superoperator is then:
\begin{equation}
    \mathscr{L}[\rho]=-2\pi i [H,\rho] + \mathscr{D}[\rho]
    \label{eq:lind}
\end{equation}
where $\mathscr{D}[\rho]$ is the dissipation superoperator acting as:
\begin{equation}
    \mathscr{D}[\rho] = \sum_{i=1}^N \frac{1}{T_1}\left(\sigma_i^+ \rho \sigma_i^- - \frac{1}{2}\left\{\sigma_i^-\sigma_i^+,\rho\right\}\right) + \sum_{i=1}^N\frac{1}{2T_\varphi}\left(Z_i \rho Z_i - \rho\right)
\end{equation}
where $\sigma_i^{\pm} = \left(X_i \pm i Y_i\right)/2$ are the raising and lowering ladder operators on qubit $i$ and $\{\cdot,\cdot\}$ the anti-commutator. We integrate Eq.~\eqref{eq:lind} by vectorizing the density matrix in column-major order $\rho \mapsto \mathrm{vec}(\rho)$, under which $H \rho \mapsto (\mathds{1}\otimes H)\mathrm{vec}(\rho)$ and $\rho H \mapsto (H^T \otimes \mathds{1})\mathrm{vec}(\rho)$ and exponentiating the resulting $4^N \times 4^N$ matrix once per drive setting. Since we simulate $N=6$, we can exponentiate without the need to use truncation or Trotterization and introduce Hamiltonian simulation errors. 

Throughout this work we hold $T_1$ at $100$~ms, comparable to the relaxation times measured in $^{28}$Si/SiGe spin qubits~\cite{amitonov_persistence_2026}. Spin relaxation is far slower than dephasing, and it could affect the reservoir only within the few input steps that the reservoir remembers. One step lasts $2\tau$ or $3\tau$, i.e., 1 or 1.5~$\mu$s, so a qubit relaxes with probability below $2\times10^{-5}$ per step. This is more than two orders of magnitude below the reset probability $\epsilon=0.02$ of Eq.~\eqref{eq:evo}, which already drives the chain toward $\ket{\mathbf{0}}$, the same state that relaxation favors. We sweep $T_2$ over $\{1,5,10,50,100,500,1000,5000\}$~$\mu$s together with the unitary reference (effectively $T_1,T_2 \rightarrow \infty$). The MG sweeps omit 500 and 1000~$\mu$s. Over this range $T_\varphi$ and $T_2$ differ by less than $3\%$, so the sweep is in effect a sweep of the pure-dephasing time. The $10$~$\mu$s column is the lower-end coherence that this type of platform typically reaches in \textit{free evolution}, and it is the column we refer to whenever we mention a \textit{realistic operating point}. The noise model and $T_\varphi$ depend on how the device is operated and differ from device to device. For example, $T_\varphi$ differs between an isolated qubit and qubits coupled by exchange, and between free and driven evolution. Our reservoir is driven with the exchange switched on, so $10~\mu$s is a free-evolution reference rather than a measured dephasing time for this operating mode.

\subsubsection{Input-driven protocol}

The input sequence $u_{1:L}$ is normalized to $[0,1]$ and drives the reservoir by interpolating between two distinct evolutions. We define two drive settings, $\mathscr{T}_0$ and $\mathscr{T}_1$, that differ in exchange strength, drive phase, transverse amplitude and duration. $\mathscr{T}_0$ evolves under Eq.~\eqref{eq:ham} with $J_0 = 2J_{ij}$, phase $\phi=0$ and $\mathbf{h}^{(0)}=(1.0,0.95,1.1,0.9,1.05,1.0)$~MHz for a time $2\tau$ with $\tau = 500~$ns. $\mathscr{T}_1$ uses couplings $J_{ij}$, phase $\phi=\pi/4$ and $\mathbf{h}^{(1)}=(1.1,1.0,0.95,1.05,1.0,0.9)$~MHz for a time $3\tau$. These amplitudes are arbitrary values within $\pm10\%$ of $1$~MHz rather than values from a specific device. Both vectors are kept fixed across reservoir realizations; only the exchange couplings change from one realization to the next. Both drive settings correspond to CPTP maps. That is, the two settings correspond to different Hamiltonians and runtimes. We should mention that although we drive the system in an analog fashion, one can always use combinations of quantum gates to drive it digitally~\cite{chen_temporal_2020}. In that case, the two maps would most probably differ also in their noise profile, since each gate has its own characteristic noise channel. The state then evolves as:

\begin{equation}
    \rho_{k+1}=(1-\epsilon)\Big[u_k \mathscr{T}_0[\rho_k]+(1-u_k)\mathscr{T}_1[\rho_k]\Big]+\epsilon \,\xi
    \label{eq:evo}
\end{equation}
\\
where $\epsilon$ is the reset rate to a $\xi$ arbitrary fixed state, both tunable hyperparameters. Here, we take $\epsilon=0.02$ (i.e., $2\%$ chance to force-restart) and $\xi = \ket{\mathbf{0}}\bra{\mathbf{0}}$, the spin-down state $\ket{\downarrow_z}^{\otimes N}$ into which spin qubits are initialized in experiment~\cite{amitonov_persistence_2026};
in principle it can be any state $\ket{\psi_0}\in\mathscr{H}^{\otimes N}$ that is easily prepared, where $\mathscr{H}$ denotes the single qubit Hilbert space. Since the above evolution is a convex combination of CPTP maps, it is itself a CPTP map for every $u_k\in[0,1]$ and the $\epsilon$ term makes it a contraction. This makes the reservoir have a fading memory and the state at step $k$ depends on the whole input history with exponentially decaying weight. All of the above claims were proven in ref.~\cite{chen_temporal_2020}.

\subsection{Benchmark tasks}
\label{sec:benchmark}

\subsubsection{\texorpdfstring{NARMA-$n$}{NARMA-n}}

The driving signal is an i.i.d. sequence drawn from a uniform distribution $u_k\sim \mathcal{U}(0,0.5)$, and the target obeys a NARMA-$n$ timeseries:

\begin{equation}
\label{eq:narma}
    y_{k+1} = \tanh\left[\alpha y_k + \beta y_k \sum_{j=0}^{n-1} y_{k-j} + \gamma u_{k-n+1}u_k + \delta\right]
\end{equation}
\\
with $(\alpha,\beta,\gamma,\delta)=(0.3,0.05,1.5,0.1)$, $y_0=0.16$, and $n=10,15,20$. The readout at step $k$ is fitted to $y_k$, and the input enters the recurrence for $y_k$ only through the single product $u_{k-n}u_{k-1}$. This is why a quadratic polynomial in the last $n+1$ taps quickly reduces the error. Moreover, the autoregressive term couples $y_k$ to a moving sum over the previous $n$ values, so the memory the task demands grows with $n$, with its nonlinearity remaining unchanged.

Although the $\tanh$ function is not usually part of the original NARMA definition, we apply it to avoid divergent feedback. Specifically, our drive is uniform on $[0,0.5]$, which is wider than the $[0,0.2]$ used with those coefficients in ref.~\cite{fujii_harnessing_2017}. At the narrower range the recurrence is bounded at every order. In contrast, at ours the $\beta y_k \sum_j y_{k-j}$ term diverges within a few hundred steps for $n=15$ and $n=20$, and the saturation is what keeps the higher orders finite. Although it is not required for stability, we also employ it at $n=10$ in order to keep a coherent definition across the three. In this way, the difference between $n=10$ and $n=20$ is a difference in memory demand rather than in recurrence. Moreover, we highlight that since we have a wider drive range than previous literature and to compensate we need a $\tanh$ function, the claims we make for NARMA are purely internal. Specifically, we compare one readout against another at matched feature count, or one coherence value against another on the same series with the same target. We flag it because a reader scanning for a familiar figure would be misled. We also note that ref.~\cite{fujii_harnessing_2017} drives its main NARMA demonstration with a product of three sinusoids. The same work uses random input, which carries no temporal structure of its own, for its memory and parity-check tasks, and repeats NARMA with white noise in an appendix. We use random input throughout, which we think is appropriate for a study on memory.

Since we need a way of checking how much input memory NARMA-$n$ actually demands independently of any reservoir, we use the delay-embedding of next-generation reservoir computing~\cite{gauthier_next_2021, bollt_explaining_2021}. We fit the same ridge readout of Eq.~\eqref{eq:output} to $y_k$ under the same protocol, but on polynomial features built directly from the last $\kappa$ inputs:

\begin{equation}
\label{eq:tap}
    \mathbf{\Psi}_{\kappa,d}(k) = \mathbf{\Phi}_d(u_k, u_{k-1},\ldots,u_{k-\kappa+1}),
\end{equation}
\\
which is independent of the reservoir employed. Since the input enters the recurrence for $y_k$ as the product $u_{k-n}u_{k-1}$, the feature set $\mathbf{\Psi}_{\kappa,2}$ contains that monomial when $\kappa \geq n+1$. The prediction error as a function of $\kappa$ therefore provides an empirical measure of the input history length needed by this baseline. We use it in Sec.~\ref{sec:narma-memory} as a memory demand indicator.

\subsubsection{Mackey--Glass}

The MG differential equation is defined as follows:

\begin{equation}
    \frac{\mathrm{d} y(t)}{\mathrm{d} t}=\frac{0.2 \,y(t-\tau_{MG})}{1+y(t-\tau_{MG})^{10}}-0.1 \,y(t)
\end{equation}
\\
and is integrated by explicit Euler at a step of 1/10 and subsampled by 10, so that one output step equals one unit of continuous time and the delay in output units is $\tau_{MG}$. We take $\tau_{MG}=16$, where the attractor is a periodic orbit, and $\tau_{MG}=17$, past the bifurcation, where it is chaotic. The first $10^3$ output steps are discarded as transient before any normalization. The delay embedding used in Fig.~\ref{fig:mg_sna} takes a lag of 15 steps, close to the delay $\tau_{MG}$.

\subsubsection{Strange non-chaotic attractor}

The SNA is generated by the noise-driven map:

\begin{equation}
\label{eq:sna}
    y_{k+1}=y_k + \frac{\sigma}{t}\left(1-y_k-\mu y_{k-1}^2+g +  D \zeta_k\right),
\end{equation}
\\
with $(\sigma,\mu,g,t)=(1,4,-0.6,4/3)$ and $\zeta_k$ standard Gaussian noise of intensity $D=0.01$, discarding $5\times 10^3$ steps of transient. Written as a two dimensional system in $(y_k,y_{k-1})$ the deterministic part is close to marginal with the largest Lyapunov exponent $\lambda_{LE}\approx -0.017$. At this noise intensity the exponent remains negative $\lambda_{LE}\approx -0.005$, which makes the attractor non-chaotic. We should mention that the stochastic noise $\zeta_k\sim \mathcal{N}(0,1)$ is redrawn at every step and is not a function of the state. This makes any closed-loop model driven only by its output unable to have any access to it, which is the substance of our result in Sec.~\ref{sec:sna}.

\subsection{Readout and the cost of a quantum observable}

\label{subsec:cost}

\subsubsection{Quantum observable sets}

After each step we measure a set $\mathbb{O}$ of expectation values on $\rho_k$. Our maximal set consists of all six single-qubit populations, transverse magnetizations and the three families of two-qubit correlators:

\begin{equation}
\label{eq:omax}
    \mathbb{O}_{\mathrm{all}} \equiv \left\{\langle Z_i \rangle, \langle X_i \rangle, \langle Z_a Z_b \rangle, \langle X_a X_b \rangle, \langle Z_a X_b \rangle \right\}
\end{equation}
\\
with $i\in\{0,\ldots,N-1\}$ and $a<b$, giving a cardinality of $|\mathbb{O}_{\mathrm{all}}|=6+6+15+15+15=57$ observables in total. Our resulting vector is $\mathbf{x}_k\in\mathbb{R}^{|\mathbb{O}_{\mathrm{all}}|}$ at step $k$ and $\mathbf{x}\in \mathbb{R}^{|\mathbb{O}_{\mathrm{all}}| \times n_{\mathrm{steps}}}$ for the whole series. Moreover, we consider restricted sets in order to ask what the extra observables actually contribute. In particular, we consider $\mathbb{O}_Z\equiv \{\langle Z_i \rangle \}$ with $|\mathbb{O}_Z|=6$, $\mathbb{O}_{\mathrm{2q}}\equiv\{ \langle Z_a Z_b\rangle, \langle X_a X_b\rangle, \langle Z_{a} \,X_b\rangle\}$ with $a<b$ and $|\mathbb{O}_{\mathrm{2q}}|=45$, $\mathbb{O}_{\mathrm{1q}}\equiv\{\langle Z_i\rangle, \langle X_i\rangle\}$ with $|\mathbb{O}_{\mathrm{1q}}|=12$ and $\mathbb{O}_{\mathrm{all}}$. We can see that $\mathbb{O}_Z\subset \mathbb{O}_{\mathrm{1q}} \subset \mathbb{O}_{\mathrm{all}}$ and $\mathbb{O}_{\mathrm{2q}}\subset \mathbb{O}_{\mathrm{all}}$, but $\mathbb{O}_{\mathrm{1q}} \cap \mathbb{O}_{\mathrm{2q}}= \varnothing$.

This is where Hilbert space as a feature generator is tested. A classical recurrent network with six state variables offers six functions of the input history. In contrast, the same six spins measured locally in the $X/Z$ bases give 57 one- and two-body expectation values, which are different nonlinear functionals of $u_{1:k}$. These additional observables probe transverse coherences and two-qubit correlations that are not captured by the single-qubit $Z$ readout~\cite{kalfus_hilbert_2022}. The question we ask here is whether feature richness is contributing, or whether it stands as a proxy for having more regressors. A $Z$-only readout can always be given more regressors by raising the polynomial degree. Therefore, we compare observable sets at matched \textit{feature} count rather than at matched polynomial degree, which we set up next. 

\subsubsection{Measurement settings}

Accessing the 57 quantum observables requires a measurement overhead. Observables compatible with the same local measurement bases can be estimated simultaneously from the same shots. This makes the number of distinct settings important for optimal resource count. Each setting requires an appropriate basis configuration and shot budget. Assigning either a $Z$ or an $X$ measurement to each qubit defines a setting $\mathcal{S}:\{0,\ldots,N-1\}\rightarrow\{Z,X\}$. Brute-force search over all $\binom{64}{4}$ sets of four settings confirms that none covers $\mathbb{O}_{\mathrm{all}}$. Hence for $N=6$, we find that five is minimal:
\begin{equation}
ZZZZZZ,\quad XXXXXX,\quad ZZZZXX,\quad ZZXXZZ,\quad ZXZXZX.
\end{equation}
The first two provide the six $\langle Z_i\rangle$ and all 15 $\langle Z_aZ_b\rangle$, and the six $\langle X_i\rangle$ and all 15 $\langle X_aX_b\rangle$, respectively. The remaining three cover all 15 mixed correlators $\langle Z_aX_b\rangle$ with $a<b$.

The mixed settings can be generated directly from the binary representations of the qubit labels, without checking each pair individually. To see this, first write each label $i\in\{0,\ldots,N-1\}$ using $\lceil\log_2N\rceil$ binary digits, including leading zeros. Each digit position defines one setting: qubit $i$ is measured in $Z$ if that digit is 0 and in $X$ if it is 1. For six qubits, the labels are $000,001,010,011,100,101$, whose three digit positions produce $ZZZZXX$, $ZZXXZZ$, and $ZXZXZX$.

For any $a<b$, the most significant digit at which their binary representations differ is 0 for $a$ and 1 for $b$. The corresponding setting therefore measures $Z$ on qubit $a$ and $X$ on qubit $b$, covering $\langle Z_aX_b\rangle$. Therefore, this construction requires at most $2+\lceil\log_2N\rceil$ overall settings.

A simpler construction uses $N+1$ settings. These are the all-$Z$ and all-$X$ settings together with the $N-1$ threshold settings
\begin{equation}
\mathcal{S}_m(i)=
\begin{cases}
Z, & i<m,\\
X, & i\geq m,
\end{cases}
\qquad m=1,\ldots,N-1.
\end{equation}
These cover every required mixed correlator because, for any $a<b$, the setting $\mathcal{S}_b$ places $a$ in the $Z$ group and $b$ in the $X$ group. For six qubits, this gives seven settings, compared with five for the binary construction. For $N\geq4$, the binary construction reduces the setting count from $N+1$ to $2+\lceil \log_2 N \rceil$.

The full observable set $\mathbb{O}_{\mathrm{all}}$ can consequently be accessed using five settings, compared to one for $\mathbb{O}_Z$. The all-$Z$ setting also yields the 15 $\langle Z_a Z_b \rangle$, so a one-setting readout could use 21 observables; we did not test that set. The benefit of accessing additional observables should therefore be weighed against the platform-dependent cost of basis changes and the shots required to achieve the desired statistical precision. Polynomial features calculated classically from these expectation values require no additional measurement settings.

\subsubsection{Polynomial features and ridge regression}

The readout is a polynomial in the measured observables. We construct all monomials of total degree $ 1 \leq \mathrm{deg}\leq d$ in the components of $\mathbf{x}$ and then min--max normalize each resulting feature. Denoting this normalized feature vector by $\mathbf{\Phi}_d(\mathbf{x})$, the prediction at step $k$ is:

\begin{equation}
\label{eq:output}
    \bar{y}_k= \mathbf{\widetilde{W}}^T \begin{pmatrix}
        \mathbf{\Phi}_d(\mathbf{x}_k) \\ 1
    \end{pmatrix}
\end{equation}
\\
with $\mathbf{\widetilde{W}}=\left(\mathbf{W}, W_c\right)$ the readout weights and bias vector. Throughout this work we take $\mathrm{deg}=2$ for the full observable set $\mathbb{O}_{\mathrm{all}}$ which yields $57 + \binom{58}{2}=1710$ features. The correlators enter as $\mathrm{deg}=1$ variables, so a quadratic readout on $\mathbb{O}_{\mathrm{all}}$ contains products of two-body expectation values. For the $Z$-only comparison we use $\mathrm{deg}=7$ on the six $\langle Z_i \rangle$, which gives 1715 features and therefore approximately matches the full set. This is the comparison that separates the effect of the observable set from that of the regressor count. For the two-qubit set we use $\mathrm{deg}=2$ with $45 + \binom{46}{2} =1080$ features. For $\mathbb{O}_{\mathrm{1q}}$ we use $\mathrm{deg}=4$, which gives $\binom{16}{4} - 1 = 1819$ features.

The optimal weights $\mathbf{\widetilde{W}}^\star$ are obtained in closed form by ridge regression, minimizing:

\begin{equation}
\mathcal{L}(\mathbf{\widetilde{W}}) = \sum_{k\in\mathbb{T}} \left|y_k-\bar{y}_k\right|^2 + \alpha_r\left(\|\mathbf{W}\|_2^2+W_c^2\right)
\end{equation}
\\
over the training index set $\mathbb{T}$, and $\alpha_r>0$ is a tunable ridge penalty hyperparameter which is applied to all coefficients, including the bias. For MG and SNA, the $y_k$ in this objective is replaced by $y_{k+1}$, since $\bar{y}_k$ predicts the next sample after the input $y_k$. The NARMA fit is deterministic. For the autonomous tasks, training has two stochastic elements: the Gaussian perturbation added to the training features and the scheduled-sampling draws. For NARMA, $\alpha_r$ is selected separately in every cell by blocked five-fold cross-validation over the training window, from 19 logarithmically spaced values in $[10^{-6}, 10^3]$. The readout is then refit on the whole training window. For MG and the SNA, we use $\alpha_r=10^{-4}$ throughout, apart from the regularization check on the chaotic task. 

\subsection{Training and evaluation methodology}

\subsubsection{Two ways of driving the reservoir}

The three tasks differ not only in the target timeseries, but also in the way the reservoir is driven and trained.

For NARMA the drive is an \textit{external} signal. We inject the i.i.d. sequence $u_{1:L}$, collect the observables, and fit the readout to the target $y_{1:L}$ that is a nonlinear filter of that same sequence. The input and target are two different objects, the task is function approximation on a driven system, and the fit is a single ridge solve.

For the MG and SNA the drive is the \textit{signal itself}, and the readout output $\bar{y}_k$ predicts the next value $y_{k+1}$. The training involves \textit{teacher forcing}, which drives the reservoir with the true $y_k$. At evaluation, the loop is closed and the reservoir is driven with its own (clamped) previous output, $u_{k+1}=\max(0,\min(\bar{y}_k,1))$, with no external signal. The raw prediction used to calculate the error remains unclamped. This is a strictly harder problem than the NARMA, since the model must have the target trajectory as an attractor of the closed-loop map it defines instead of being only a function approximator. We call the NARMA setting \textit{open-loop} and this one \textit{autonomous}. It is worth highlighting that the teacher-forced phase is open-loop in the same sense, although drive is the target instead of an independent input. Nevertheless, their training method is different from the one we show here.

Training and evaluation windows are specified separately for each task. Using zero-based indices and half-open intervals, the NARMA experiments discard the first 100 samples as washout, fit the readout over $[100,4250)$, and evaluate over $[4250,5000)$. For MG, the first 200 samples are used for washout, the training window is $[200,10000)$, and the evaluation window is $[10000,12000)$. The first 300 evaluation steps use teacher forcing, followed by 1700 autonomous updates. Of the resulting predictions, 1699 have available next-step targets and enter the evaluation. For both autonomous tasks the reservoir is reinitialized to $\xi$ at the start of the evaluation window, so the teacher-forced steps that open it also serve as washout.

For SNA, the first 200 samples are used for reservoir washout, and the training window is the same as in MG. The evaluation window is $[12000,22000)$. The first 1000 evaluation steps use teacher forcing, followed by 9000 autonomous updates. All resulting autonomous predictions have available next-step targets, corresponding to target indices $[13001,22001)$, so that $K_{\mathrm{eval}}=9000$. The retained readout is selected by the lowest autonomous NMSE on this window. For the autonomous tasks, the evaluation window is also used for model selection. We evaluate the teacher-forced bootstrap and each subsequent DAgger iteration on this window, selecting the readout with the lowest NMSE or the longest valid prediction time $\mathcal{V}$, depending on the task. The regression coefficients are fitted using training-window samples, while the retained iteration is determined by the evaluation-window targets. We report the selected readout's performance on that same window, without an additional independent test window. The reported scores may therefore be optimistic relative to the performance on independent data. Every configuration goes through the same selection protocol, so comparisons between coherence values and between reservoirs share this optimism, although not necessarily to the same degree.

\subsubsection{Why teacher forcing is not sufficient}

The standard method fits Eq.~\eqref{eq:output} on teacher-forced states, which are states visited when the true signal is fed in at every step. Although it minimizes the one-step error during the training phase, it does not say anything about what happens when the loop is closed. In general, the trained map is a deterministic dynamical system and the target orbit is not an attractor of it. As a consequence, the closed loop either contracts onto a fixed point or departs from the target trajectory. In order to confirm that this is not a matter of insufficient regularization, we swept the ridge parameter over $\alpha_r\in\{10^{-6}, 10^{-4}, 10^{-2},1\}$. On the chaotic target, it left the ordering of every coherence comparison unchanged over all 12 combinations of $\alpha_r$ and training seed. The selection did not keep the teacher-forced bootstrap for any value of $\alpha_r$ and the aggregated fit was selected in all 24 runs.

This failure mode is the exposure bias of autoregressive models~\cite{ranzato_sequence_2016, he_exposure_2021}, and is not specific to quantum reservoirs. The model is trained only on states drawn from the data distribution and then asked to operate on states drawn from its own. As a result, it drifts away from the moment a small error accumulates.

\subsubsection{In-loop training with dataset aggregation}

In order to avoid the problem mentioned above we therefore have to train the readout on the states it will actually visit. We show the pipeline in Algorithm~\ref{alg:dagger}. We start from the teacher-forced bootstrap fit $\mathbf{\widetilde{W}}^{(0)}$. At each iteration $q$ we run the reservoir over the training window feeding back the model's own prediction with probability $p_q$ and the true value with probability $1-p_q$. This routine is called \textit{scheduled sampling}~\cite{bengio_scheduled_2015}. We label every state on that run, including off-attractor ones, with the \textit{true} next value. We append those pairs to the aggregated dataset $\mathbb{D}_q=\mathbb{D}_{q-1}\cup \{(\mathbf{x},y)\}$ and refit $\mathbf{\widetilde{W}}^{(q)}$ on the whole. This is called \textit{dataset aggregation (DAgger)}~\cite{ross_reduction_2011} and we borrow it from imitation learning. The aggregated dataset $\mathbb{D}_q$ teaches the readout both what next step value is on the attractor and also how to get back to the attractor from the neighborhood it drifts into.

\begin{algorithm}[t]
\caption{In-loop readout training. The reservoir map $\mathscr{T}$ and the set of polynomial monomials are fixed, while $\mathbf{\widetilde{W}}$ is learned.}
\label{alg:dagger}
\begin{algorithmic}[1]
\Require target $y_{1:L}$, windows $[k_0,k_1)$ for training and $[e_0,e_1)$ for evaluation, number of iterations $Q$, final feedback probability $p_{\textrm{end}}$, ridge parameter $\alpha_r$, feature noise scale $\varsigma$ 
\State $\mathbb{D} \gets \varnothing$; \; $\mathbf{\widetilde{W}} \gets \varnothing$
\For{$q=0, 1, \ldots, Q$}
    \State $p_q \gets 0$ if $q=0$ else $p_{\textrm{end}}\, q / Q$
        \Comment{scheduled sampling ramp} \label{alg:sched}
    \State $\rho \gets \xi$; \; $\bar{y} \gets \varnothing$
    \For{$k = 0, \ldots, k_1 - 1$}
    \State $u_k \gets \max(0, \min(\bar{y},1))$ with probability $p_q$ if $k \geq k_0$ and
           $\mathbf{\widetilde{W}} \neq \varnothing$, else $u_k \gets y_k$
           \Comment{drive from own output or from truth}
    \State $\rho \gets \mathscr{T}[\rho; u_k]$ by Eq.~\eqref{eq:evo};
           \; $\mathbf{x} \gets$ observables of $\rho$
    \If{$k \geq k_0$}
        \State $\mathbb{D}\gets \mathbb{D}\cup \{(\mathbf{x},\, y_{k+1})\}$
            \Comment{label with the TRUE next value} \label{alg:label}
        \EndIf
        \State $\bar{y} \gets \mathbf{\widetilde{W}}^{T}\begin{pmatrix}
            \mathbf{\Phi}_d(\mathbf{x}) \\ 1
        \end{pmatrix}$ \textbf{if} $\mathbf{\widetilde{W}} \neq \varnothing$
        \EndFor
        \State recompute the min--max constants over all $\mathbb{D}$
            \Comment{normalization tracks the visited distribution}
        \State $\mathbf{\widetilde{W}}^{(q)} \gets \underset{\mathbf{\widetilde{W}}}{\mathrm{argmin}}\, \mathcal{L}(\mathbf{\widetilde{W}})$
            on $\mathbb{D}$, features perturbed by $\mathcal{N}(0,\varsigma^2)$
        \State $\mathcal{M}_q \gets$ autonomous score of $\mathbf{\widetilde{W}}^{(q)}$ on $[e_0, e_1)$ \Comment{closed loop, not one-step}
        \State $\mathbf{\widetilde{W}} \gets \mathbf{\widetilde{W}}^{(q)}$
    \EndFor
    \State \Return $\mathbf{\widetilde{W}}^{(q^\star)}$ with
    $q^\star =
    \begin{cases}
        \underset{q}{\mathrm{argmin}}\,\mathcal{M}_q, \quad  \mathrm{if} \,\,\, \mathcal{M}_q = \mathrm{NMSE} \\[.4cm]
        \underset{q}{\mathrm{argmax}}\,\mathcal{M}_q, \quad \mathrm{if} \,\,\, \mathcal{M}_q = \mathcal{V}
    \end{cases}$
\end{algorithmic}
\end{algorithm}
Each fitted readout is stored with its corresponding min--max normalization constants, which are retained with the selected iteration. We ramp the feedback probability linearly as $p_q = p_{\textrm{end}}\, q/Q$ with $Q=15$ iterations and $p_{\textrm{end}}=0.8$, both of which are hyperparameters. We also add a small Gaussian perturbation of standard deviation $\varsigma=10^{-3}$ to the normalized training features at every refit. The performance of the closed-loop map is not monotonic in $q$ and the procedure usually passes through a rough transient. The best readout is some $0\leq q^\star\leq Q$, so we evaluate the autonomous performance after each iteration and keep the best snapshot:

\begin{equation}
\begin{aligned}
    q^\star &= \begin{cases}
        \underset{0\,\leq q\, \leq Q}{\mathrm{argmin}}\,\mathcal{M}_q, \quad  \mathrm{if} \,\,\, \mathcal{M}_q = \mathrm{NMSE}, \\[.4cm]
        \underset{0\,\leq q\, \leq Q}{\mathrm{argmax}}\,\mathcal{M}_q, \quad \mathrm{if} \,\,\, \mathcal{M}_q = \mathcal{V},
    \end{cases} \\[.4cm]
    \mathbf{\widetilde{W}}^\star &= \mathbf{\widetilde{W}}^{(q^\star)}
\end{aligned}
\end{equation}
\\
where $\mathcal{M}$ is the selection metric defined in Sec.~\ref{sec:metrics}. Given that the bootstrap fit $\mathbf{\widetilde{W}}^{(0)}$ is included in the comparison, the procedure can in principle fall back on pure teacher forcing (i.e., baseline) whenever the in-loop iterations fail to improve on it.

We note that across all 70 non-chaotic runs of the coherence sweep the bootstrap fit was never selected. The median choice was iteration 11 out of 15, while 53 out of 70 runs picked an interior iteration $0<q^\star<Q$ and the remaining 17 selected $q^\star=Q$. The aggregated fit therefore scored better than the bootstrap on the selection window in every run. Finally, in about 1/4 of our runs it kept improving until $Q$, meaning it could potentially offer improvements if we extended the iteration range.

\subsubsection{Metrics}
\label{sec:metrics}

For evaluation, predictions are indexed by their target time: $\widehat{y}_k=\bar{y}_k$ for NARMA and $\widehat{y}_k=\bar{y}_{k-1}$ for MG and SNA. The metrics below compare $\widehat{y}_k$ with $y_k$.

For the NARMA, non-chaotic MG and SNA tasks, we score the evaluation window with the normalized mean-square error (NMSE), which is also the selection metric for the two autonomous ones:

\begin{equation}
    \mathrm{NMSE} = \frac{\sum_{k} (\widehat{y}_k-y_k)^2}{\sum_k (y_k - \mu_y)^2} , \quad \mu_y = \frac{1}{|\mathscr{E}|}\sum_{k\in\mathscr{E}}y_k
\end{equation}
\\
with $\mathscr{E}$ the evaluation index set, so that NMSE $=1$ is the performance of predicting the mean and values above one are worse than the constant-mean baseline. 

For the chaotic target we select and report by the \textit{valid prediction time} $\mathcal{V}$, the standard horizon metric in the chaotic RC literature~\cite{pathak_model-free_2018, vlachas_backpropagation_2020}. Two trajectories that start arbitrarily close separate exponentially, so the NMSE over a long autonomous window saturates regardless of the model's quality, and we quote it only as a secondary measure.
It is defined as:
\begin{equation}
\label{eq:vtime}
\mathcal{V}=\min\left(
    \left\{
    \ell\in\{0,\ldots,K_{\mathrm{eval}}-1\} :
    |\widehat{y}_{k_a+\ell}-y_{k_a+\ell}|>\theta\,\sigma_y
    \right\}
    \cup\{K_{\mathrm{eval}}\}
\right)
\end{equation}
where $k_a$ is the target-time index of the first scored autonomous prediction, $\sigma_y$ is the standard deviation of the aligned target values over the scored autonomous window, and $\theta$ a threshold which we set to $\theta=1$ here. We see that $\mathcal{V}$ effectively calculates the time before the error exceeds one standard deviation of the signal. Here, $K_{\mathrm{eval}}$ is the number of autonomous predictions for which aligned target values are available, and $\ell$ counts steps from the first scored autonomous prediction. If the error never exceeds the threshold within this window, we set $\mathcal{V}=K_{\mathrm{eval}}$. Thus, $0\leq\mathcal{V}\leq K_{\mathrm{eval}}$, with $K_{\mathrm{eval}}=1699$ in our MG evaluations. Reaching this ceiling means that the prediction remains within the chosen error threshold throughout the scored autonomous window, and its validity beyond that window is not assessed. Unlike ref.~\cite{vlachas_backpropagation_2020}, we do not normalize $\mathcal{V}$ by the maximum Lyapunov exponent since we compare a single system against itself across coherence values instead of different attractors. This makes the raw step count a more directly interpretable quantity.

\subsection{Use of AI tools}
We used a large language model to refine the language of the manuscript, which included rewording and condensing the authors' text, and to produce the TikZ source for the schematic in Fig.~\ref{fig:scheme} to the authors' design. The authors directed the work, reviewed all text, and take full responsibility for the content of this article.

\subsection{Code availability}
The simulation and analysis code used in this study is available from the corresponding author on reasonable request subject to approval by Equal1 Labs.

\backmatter

\bmhead{Acknowledgements}
We would like to thank our colleagues Andrii Semenov for insightful discussions and Niall Murphy for providing assistance with the code. Part of this work appeared in the PhD thesis of N.P.~\cite{petropoulos_thesis}.

\bmhead{Author Contributions}
Nikolaos Petropoulos: Conceptualization, Methodology, Software, Writing -- original draft. Elena Blokhina: Conceptualization, Writing -- review \& editing, Supervision, Project administration, Funding acquisition.

\bmhead{Data availability}
The datasets generated and analyzed during the current study are available from the corresponding author on reasonable request.

\bmhead{Competing Interests}
Both authors are affiliated with Equal1 Labs, a company developing silicon spin-qubit quantum computers. The authors declare no other competing interests.

\noindent

\bibliography{bibliography_1, bibliography_2}


\end{document}